\documentclass[aip,pop,reprint]{revtex4-1}
\usepackage{graphicx}
\usepackage{amsmath}
\usepackage{amssymb}
\usepackage{amscd}
\usepackage{bbm}
\usepackage{epstopdf}
\usepackage{color}

\usepackage{listings}

\usepackage{tikz,pgfplots}
\pgfplotsset{compat=newest} 
\pgfplotsset{plot coordinates/math parser=false}

\usepackage[colorlinks=true,
		citecolor=blue,
		linkcolor=blue,
		anchorcolor=blue,
		filecolor=blue,
		menucolor=blue,
		runcolor=blue,
		urlcolor=blue,
		unicode
		]{hyperref}
\usepackage[all]{hypcap}

\newcommand{\mycomment}[1]{}

\newcommand{\curl}{\nabla \times}

\newcommand{\B}{{\bf{B}}}

\newcommand{\J}{{\bf J}}

\newcommand{\V}{{\bf v}}

\newcommand{\Omg}{{\pmb \Omega}}

\begin{document}

\title{
Propagation speed, linear stability, and ion acceleration in radially imploding Hall-driven electron-magnetohydrodynamic shocks
}

\author{A.~S.~Richardson}
\author{S.~B.~Swanekamp}
\author{S.~L.~Jackson}
\affiliation{Plasma Physics Division, Naval Research Laboratory, Washington, DC 20375, USA}
\author{D.~Mosher}
\author{P.~F.~Ottinger}
\affiliation{Independent consultant through Syntek Technologies, Arlington, VA 22203}

\date{\today}

\begin{abstract}
Plasma density gradients are known to drive magnetic shocks in electron-magnetohydrodynamics (EMHD). Previous slab modeling has been extended to cylindrical modeling of radially imploding shocks. The main new effect of the cylindrical geometry is found to be a radial dependence in the speed of shock propagation. This is shown here analytically and in numerical simulations. Ion acceleration by the magnetic shock is shown to possibly become substantial, especially in the peaked structures that develop in the shock because of electron inertia.
\footnote{This work was supported by the Naval Research Laboratory Base Program.}
\end{abstract}

\maketitle


There are many interesting phenomena that arise in electron-magnetohydrodynamics (EMHD), which is a model of plasma where the ions are taken to be fixed and electrons are modeled as a fluid.\cite{Gordeev1994215} One such phenomenon is that the interaction of a current channel with a background-plasma density gradient can drive magnetic shock waves.\cite{Kingsep_EMHD,Kingsep1983} In a one-dimensional Cartesian reduction, these shock waves are described by Burgers' equation, which has analytical shock solutions with a hyperbolic tangent form. Modifications to this solution are necessary if the width $D$ of the shock front is comparable to the electron inertial length $\delta_e = c/\omega_{pe}$.\cite{kalda:small-scale,Gordeev1994215} Additionally, it was recently shown that electron-inertial effects can, in two dimensions, cause the shock front to go unstable and generate magnetic vortices.\cite{richardson:2016:EMHD} That work was done in the context of a planar shock wave in Cartesian geometry. In this paper, we demonstrate that the previous results also hold for a cylindrical shock wave, such as might be encountered in the geometry of z-pinch, imploding liner, or plasma-filled diode (PFD) experiments.\cite{doi:10.1063/1.354711,1347255,kovalchuk_zherlitsyn_pedin_2010} EMHD physics has been hypothesized to be applicable to PFD experiments\cite{Gordeev1994215}, and future research could compare the instability reported in this paper to experimental results.
The main difference between a cylindrical and a planar EMHD magnetic shock is that the planar shock moves at constant speed, while the speed of the cylindrical shock depends on its radial location. In the following sections, this dependence is shown analytically and in numerical simulations, and the implications for possible ion acceleration are discussed. Additionally, we show linear stability calculations and two-dimensional simulation results which demonstrate that electron inertia effects cause the magnetic front to break up into vortices in cylindrical shocks, as was previously shown in planar shocks. More details about these calculations can be found in Ref.~\onlinecite{richardson:2016:EMHD}, where similar calculations were done in Cartesian geometry.

In the EMHD model, a fluid approximation is used to describe the electrons while the ions are taken to be fixed. Combining the curl of the electron momentum equation with Maxwell's laws gives a set of dynamical equations for the magnetic field and the vorticity $\Omg$ of the canonical electron momentum\cite{Gordeev1994215}:
\begin{eqnarray}
\partial_t {\pmb \Omega} &=& \curl (\V \times {\pmb \Omega}) - m c \nu \curl \V / e, \\
{\pmb \Omega} &=& \curl (mc\V/e) - \B,
\end{eqnarray}
where $\nu$ is the electron-ion collision frequency. The electron fluid velocity $\V$ is related to the magnetic field through $\V = -\frac{c}{4\pi ne} \nabla \times \B$, because the electrons carry the current, i.e. $\J = -en\V$. These equations are simplified by performing the following normalizations: $\B/B_0\rightarrow\B$, $\Omg/B_0\rightarrow\Omg$, ${\bf x}/\delta_e\rightarrow{\bf x}$, $t \omega_{ce}\rightarrow t$, and $n/n_0\rightarrow n$. With these normalized units, the equations of EMHD are:
\begin{eqnarray}
\partial_t {\pmb \Omega} &=& \curl (\V \times {\pmb \Omega}) - \nu \curl \V , \label{eq:vort} \\
{\pmb \Omega} &=& \curl \V - \B \label{vort-def}, \\
\V &=& -\frac{1}{n} \curl \B.
\end{eqnarray}
 In the previous slab geometry, the magnetic field is in the $y$ direction, the background plasma density gradient is in the $z$ direction, and the Hall-driven magnetic shock travels in the $x$ direction. In this paper, the magnetic field is in the $\theta$ direction, the density gradient is again in the $z$ direction, and the shock propagates in the $r$ direction. We will consider the imploding case, where the shock travels from large radius inward toward the axis. The EMHD equations further simplify when it is assumed that the system is invariant in the direction of the magnetic field.

Note that the cylindrical geometry considered in this paper was also discussed in Section 6 of Ref.~\onlinecite{Gordeev1994215}. However, in that work, they did not examine radially imploding magnetic shocks, which are the subject of this paper.

Another simplification used in the previous work, and again assumed to be the case here, is to only consider the scenario where the linear density gradient is small compared to the size $L_z$ of the system in the $z$ direction. In this case, we can write the normalized density as $n = 1+ z/L_n \simeq 1$ and we have the smallness parameter $L_z\nabla n = L_z/L_n \ll 1$. Then $\Lambda \equiv \partial_z \ln n = L_n^{-1} (1+z/L_n)^{-1} \simeq L_n^{-1}$. In this approximation, the effect of the density gradient comes through the terms in the EMHD equations that are proportional to $\Lambda$, which is constant. 

With these simplifications, the EMHD equations become a pair of scalar equations for the magnitudes $B$ and $\Omega$ of the magnetic field and the vorticity:
\begin{eqnarray}
\Omega &=&  \partial_r \frac{1}{r} \partial_r (r B) + \partial_z^2 B - { \Lambda} \partial_z B - B, \label{2D-EMHD-1}\\
\partial_t { \Omega} &=& \left\{ B, \Omega \right\} - \frac{\Lambda \Omega}{r} \partial_r (rB) -\nu (\Omega + B) ,\label{2D-EMHD-2}
\end{eqnarray}
where we have used the notation
\begin{eqnarray}
\left\{ B, \Omega \right\} = \frac{1}{r}\partial_r(rB) \partial_z \Omega - \partial_r \Omega \partial_z B .
\end{eqnarray}
Equations \eqref{2D-EMHD-1} and \eqref{2D-EMHD-2} are the subject of this paper. 

In order to better understand the shock-wave solutions to these equations, first consider the case where the solutions are assumed to be invariant in the $z$ direction. The governing equations then only depend on $r$, but are still a relatively complex set of nonlinear equations:\\ \\
{\em 1D cylindrical EMHD shock model}
\begin{eqnarray}
\Omega &=& \partial_r \frac{1}{r} \partial_r (r B)  - B \label{cyl1}\\
\partial_t { \Omega} &=&  -\frac{\Lambda \Omega}{r}\partial_r(rB)  -\nu (\Omega + B) \label{cyl2}
\end{eqnarray}
One further simplification is to drop the terms in these equations that arise from the electron inertia, as was done in Ref.~\onlinecite{richardson:2016:EMHD}. In this case, Eq.~\eqref{cyl1} becomes $\Omega = -B$ to lowest order. While it seems that this would allow the last term in Eq.~\eqref{cyl2} to be dropped, this is not the case if $\nu$ is large enough. Since we are not making any assumptions about the size of $\nu$ at this point, that term will be retained.
The result is an equation that looks similar to the Burgers' equation that arises in the EMHD slab model, but with some variations because of the differences in the radial derivatives:\\ \\
{\em 1D cylindrical EMHD shock model (neglecting electron inertia)}
\begin{eqnarray}
\partial_t { B} &=&  -\frac{\Lambda B}{r}\partial_r(rB)  +\nu \partial_r \frac{1}{r} \partial_r (r B)  \label{1D-no-inertia}
\end{eqnarray}
We can estimate the speed of the imploding shock wave by multiplying Eq.~\eqref{1D-no-inertia} by $r^2$ and integrating over $r$. This gives
\begin{eqnarray}\label{shock_int1}
\partial_t \int_0^\infty r^2 B \, dr &=& -\frac{\Lambda}{2} \left[(rB)^2\right]_0^\infty - 2\nu \left[rB\right]_0^\infty \\
&=& -\frac{\Lambda}{2} u_0^2 - 2\nu u_0. \notag
\end{eqnarray}
Now consider the case where $rB$ has the form of a shock wave collapsing towards the axis, with $rB=u_0={\rm constant}$ for $r>r_s(t)$, and $rB=0$ for $r\leq r_s(t)$, where $r_s(t)$ is the location of the shock front. For this shock wave, the left-hand side of Eq.~\ref{shock_int1} becomes
\begin{eqnarray}\label{shock_int2}
\partial_t \int_0^\infty r^2 B \, dr = \partial_t \int_{r_s(t)}^\infty r u_0 \, dr 
= - \frac{dr_s}{dt} r_s(t) u_0,
\end{eqnarray}
where the last term comes from applying Leibniz's rule. Combining Eqs.~\ref{shock_int1} and \ref{shock_int2} 
implies that the shock front moves towards the axis with speed $v_{\rm shock} \equiv \vert dr_s/dt\vert$:
\begin{eqnarray}
v_{\rm shock} = \left\vert \frac{1}{2r}(\Lambda u_0 + 4\nu) \right\vert. \label{v_shock}
\end{eqnarray}
To lowest order one would expect that adding back the terms due to electron inertia will not change this shock speed. For the initial condition being considered, with $rB > 0$, this calculation also implies that $\Lambda < 0$ and $|\Lambda u_0| > 4\nu$ is needed in order for the shock to move towards the axis (i.e., $dr_s/dt < 0$).

Notice that the shock speed now explicitly depends on the radius of the shock location. This is in contrast to the EMHD slab model studied previously\cite{richardson:2016:EMHD}, where the shock speed $v_{\rm slab}$ is constant, and $v_{\rm slab} = \Lambda B_0/2$. This implies that a cylindrical EMHD shock accelerates as it implodes towards the axis, and its speed increases like $1/r$.

\begin{figure}[htbp]
(a)\hspace*{\fill}
\begin{center}
\includegraphics[]{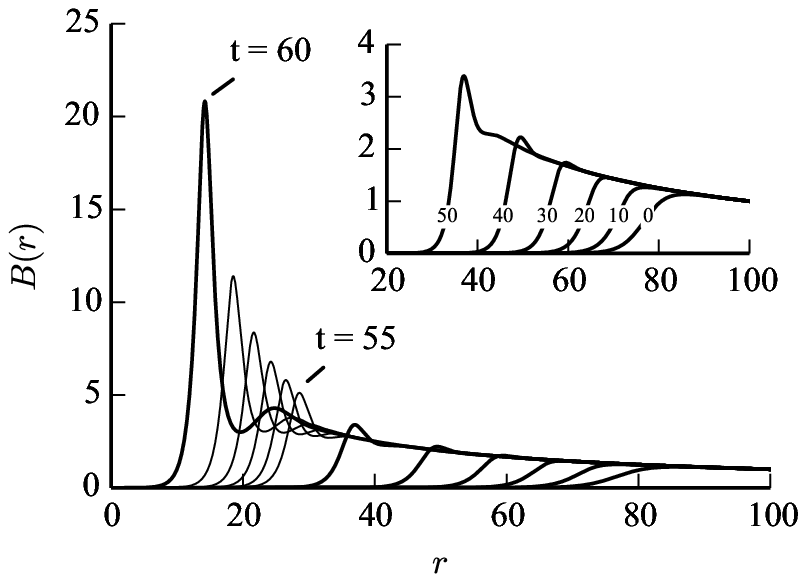}
\end{center}
(b)\hspace*{\fill}
\begin{center}
\includegraphics[]{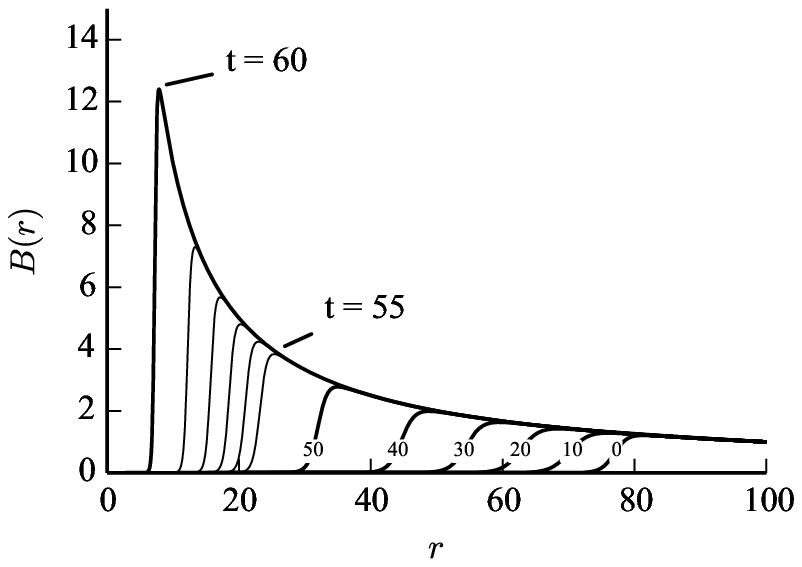}
\caption{Numerical solution to the cylindrical EMHD equations; (a) with electron inertia and (b) without electron inertia. The numbers labeling the curves are the time in normalized units, and the acceleration of the front as it approaches the axis can be clearly seen.
\label{fig:numerical}}
\end{center}
\end{figure}

In order to verify this shock speed, a one-dimensional solver was used to solve the 1D EMHD equations both with the inertial terms [Eqs.~\eqref{cyl1} and \eqref{cyl2}] and without them [Eq.~\eqref{1D-no-inertia}]. For these simulations, the parameters were $\Lambda = -1$, $\nu = 1$, and $u_0 = 100$, and the initial shape is \[rB(r) = u_0( 0.5 \tanh(  (r - 77)/5 ) +0.5 ).\] Figure \ref{fig:numerical} shows the numerical solutions. As in slab geometry model, the electron inertia terms cause the shock front to develop a peaked structure, with oscillations behind the front which can be seen by comparing Fig.~\ref{fig:numerical}(a) and Fig.~\ref{fig:numerical}(b). Note that because of these peaks, the value of $rB$ can locally become much greater than $u_0$ when the effects of electron inertia are retained. Without the inertial terms, the maximum value that $rB$ attains at any radius is $u_0$. 
Also, as predicted above, these numerical results show that the shock front accelerates as it approaches the axis.

\begin{figure}[htbp]
\begin{center}
\includegraphics[]{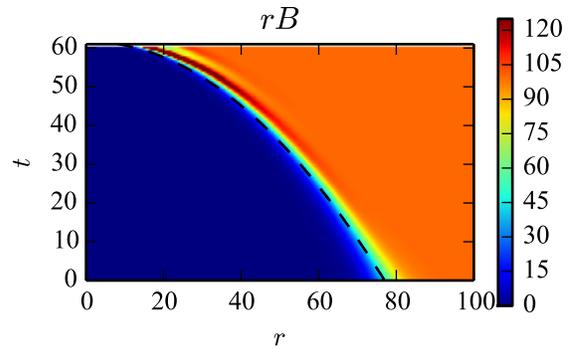}
\caption{Streak plot in $r$, $t$ plane of the numerical solution to the cylindrical EMHD equation with inertial terms. The dashed curve is obtained by integrating the velocity in Eq.~\eqref{v_shock}. Parameters: $r_0 = 77$, $u_0 = 100$, $\Lambda = -1$, and $\nu = 1$.
\label{fig:streak}}
\end{center}
\end{figure}

With the value of speed given by Eq.~\eqref{v_shock} we get the following for the analytically predicted shock front location for a shock starting at $r=r_0$:
\begin{eqnarray}
r_s(t) =  \sqrt{r_0^2 + (\Lambda u_0 + 4\nu) t}.
\end{eqnarray}
This is compared in Fig.~\ref{fig:streak} to a streak image of the numerical solution (with inertial terms) in the $r$,$t$ plane. The dashed curve gives the predicted location of the shock front, where the color contours represent the value of $rB$. The predicted shock location agrees well with the numerical solution.

In order to estimate how this EMHD shock can accelerate ions, we use a pressure balance argument used by Rosenbluth in his collisionless magnetic piston model\cite{rosenbluth_pinch}. As the magnetic shock implodes towards the axis, the magnetic pressure in the front is balanced by a kinetic pressure due to the momentum imparted to the ions. The momentum change of the ions in an annular volume $2\pi r \Delta r \Delta z$ is
\begin{eqnarray}
\Delta p &=& M (v_f - v_i) = 2\pi r \Delta r \,\Delta z\, m n v_f,
\end{eqnarray}
where $m$ is the ion mass and $v_f$ is the speed to which the ions are accelerated from $v_i=0$.
This momentum change happens during the time in which the shock passes through the region $\Delta r$, which is related to the speed of the shock: $\Delta t = \Delta r/ v_{\rm shock}(r)$. This gives a  kinetic pressure of
\begin{eqnarray}
P_k &=& \frac{{\Delta p}/{\Delta t}}{2\pi r \Delta z} =  m n v_f v_{\rm shock}(r).
\end{eqnarray}
The final velocity $v_f$ to which the ions are accelerated is found by balancing this kinetic pressure with the magnetic pressure $\frac{B^2}{8\pi}$,
which gives a ratio of ion speed to shock speed of 
\begin{eqnarray}
\frac{v_f}{v_{\rm shock}} &=& \frac{B^2}{8\pi m n v_{\rm shock}^2}.
\end{eqnarray}
Using Eq.~\eqref{v_shock} for the shock velocity, we get
\begin{eqnarray}\label{eq:v_final}
\frac{v_f}{v_{\rm shock}} &=& \frac{(2rB)^2}{8\pi m n \left( {\Lambda u_0} + 4\nu \right)^2}.
\end{eqnarray}
The numerical solutions for the case without inertia [see Fig.~\ref{fig:numerical}(b)] show that behind the shock front the solution is approximately given by $rB = u_0 =  {\rm const}$. For constant $rB$, the velocity ratio in Eq.~\eqref{eq:v_final} is also approximately constant. If this ratio is much less than one, then the ions don't accelerate much, and this model of ion acceleration is approximately valid.

\begin{figure}[htbp]
\begin{center}
\includegraphics[]{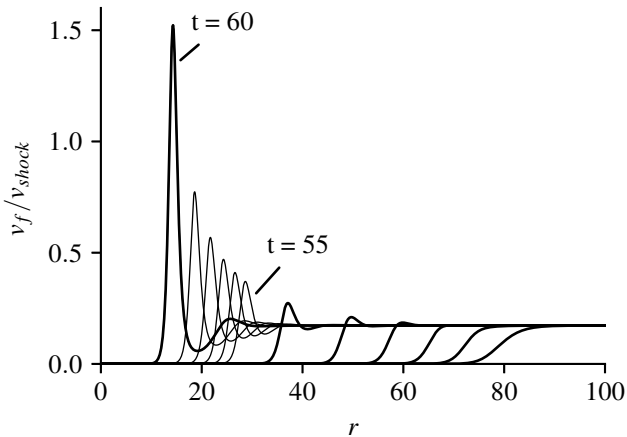}
\caption{Ratio of ion final velocity to the speed of the shock front in the case where electron inertia effects are retained. When this ratio exceeds one, the EMHD model is no longer valid, since ions are accelerated to speeds exceeding the shock front speed.
\label{v_final_ions}}
\end{center}
\end{figure}

In the case where electron inertia effects are retained [Fig.~\ref{fig:numerical}(a)], the situation is considerably different because of the structure of the shock front. In this case, the shock front develops into a peaked structure, with magnetic fields that can be significantly larger than the previous case of $B\simeq u_0/r$. As these structures develop, the final ion velocity attained due to magnetic pressure can (locally) increase to large values. This effect is illustrated in Fig.~\ref{v_final_ions}, where the velocity ratio in Eq.~\eqref{eq:v_final} is plotted for various times (with ion mass taken to be the proton mass, $m=1$). As is seen in the figure, the ratio becomes substantial and even exceeds unity near time $t=60$. When this happens, the ions are predicted to move faster than the shock front, implying that the ``stationary-ions'' assumption of the EMHD model is clearly not valid. 

Note that ion acceleration could be a more significant effect in this cylindrical implosion than in the slab model. This is because the development of peaked structures in the shock is due to electron inertia. The effect of electron inertia becomes important when the shock width is comparable to the electron skin depth. However, the shock width is determined by the balance of nonlinear steepening with collisional diffusion, and when the shock moves faster, it is narrower because diffusion has less time to broaden the front. Thus, electron-inertia effects are more important as the shock accelerates towards the axis, and so ion acceleration could also become more pronounced.

As in the previously studied slab model, there are additional effects that can come into play when variations are allowed in the $z$ direction. In particular, the magnetic shock front can be unstable to a Kelvin-Helmholtz (KH) like mode, which will break the shock front apart into vortices that propagate faster than the shock front itself is predicted to move. This effect was examined for the cylindrical model by first doing linear stability calculations of a cylindrical shock front, and then computing solutions to the full 2D equations \eqref{2D-EMHD-1} and \eqref{2D-EMHD-2}.

\begin{figure}[htbp]
\begin{center}
\includegraphics[width=7cm]{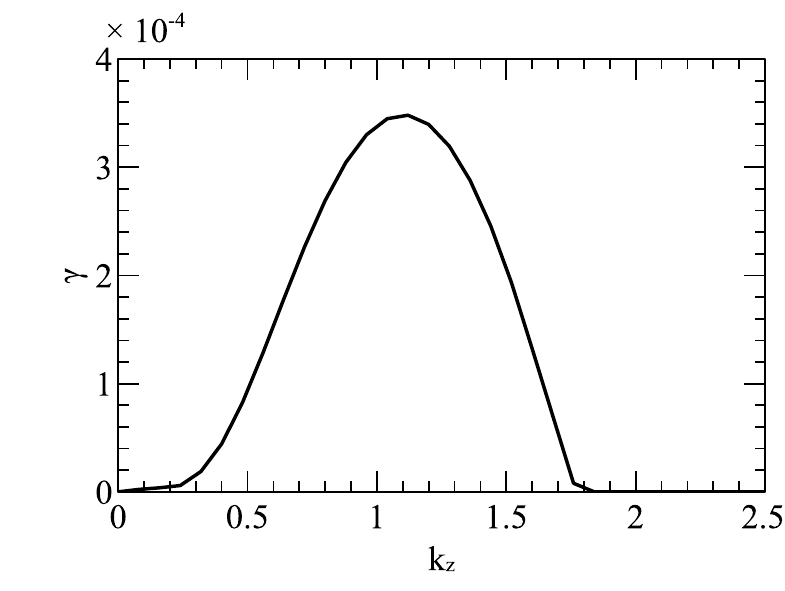}
\caption{Growth rate of the fastest growing mode as a function of wavenumber in the $z$ direction.
\label{fig:growth}}
\end{center}
\end{figure}

Linear stability calculations for a shock with a profile given by $rB(r) = u_0( 0.5 \tanh(  r - 77 ) +0.5 )$ show that there exists a range of wave numbers that are unstable. The growth rate $\gamma$ is the imaginary part of the numerically computed linear eigenvalues, and is shown in Fig.~\ref{fig:growth} as a function of wave number $k_z$. For more details about this linear analysis, see Ref.~\onlinecite{richardson:2016:EMHD}, where the calculation is described in detail for the slab model.

\begin{figure}[h]
\begin{center}
\includegraphics[width=8cm]{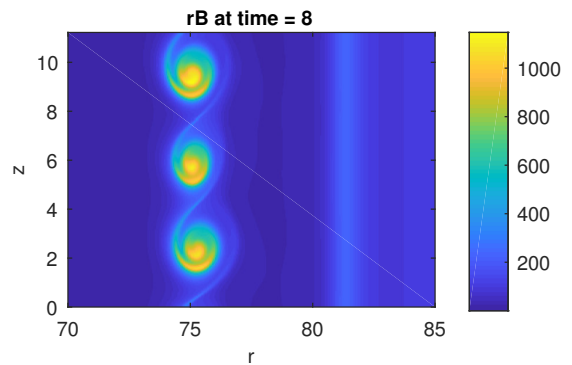}
\caption{2D fluid simulation result at time $t=8$, obtained by numerically solving Eqs.~\ref{cyl1} and \ref{cyl2}. The magnetic shock front has just gone unstable and broken apart into a series of magnetic vortices at this time.  $\Lambda = -1$ and $\nu = 0.0375$.
\label{2Dsim}}
\end{center}
\end{figure}

\begin{figure}[htbp]
\begin{center}
\includegraphics[width=8cm]{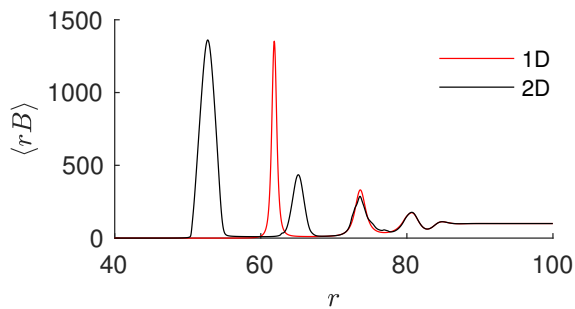}
\caption{Comparison of 1D and 2D numerical solutions at $t=21$, well after the time that the front has broken into vortices. The red curve shows the 1D result, while the black curve is the the 2D result, integrated over the $z$ direction. Note that the vortex in the 2D simulation has moved farther inward than the 1D front. $\Lambda = -1$ and $\nu = 0.0375$.
\label{1D2D}}
\end{center}
\end{figure}

In order to see the nonlinear structure that develops out of these unstable modes, two dimensional simulations were performed using the BOUT++ framework.\cite{Dudson2009}  These simulations show that the magnetic front goes unstable to a KH-like mode, just as it does in the slab model. The KH mode develops nonlinearly, breaking apart into a series of magnetic vortices. This can be seen in Fig.~\ref{2Dsim}. These vortices then travel through the plasma faster than the 1D analysis predicts. This is seen by comparing the 1D simulation to the 2D simulation averaged by integrating over the $z$ direction, as shown in Fig.~\ref{1D2D}. Also as in the slab case, electron inertia must be retained for this mode to become unstable. Note that the simple pressure-balance derivation for ion motion outlined above will no longer apply once these vortices develop. Very strong magnetic fields become entrained in the vortices, and the possibility of ion motion in the vortices should be considered. In slab geometry, it was shown that the EMHD assumption of quasi-neutrality breaks down\cite{:/content/aip/journal/pop/21/11/10.1063/1.4902101}, further complicating the physics of these vortices.

In summary, it has been shown that Hall-driven EMHD shocks in cylindrical geometry behave similarly to EMHD shocks in slab geometry. However, rather than moving at constant speed, these shocks accelerate as they implode towards the axis. This accelerating shock steepens, and the nonlinear effects of electron inertia become more important than the smoothing due to collisions. This in turn implies that ion acceleration to large speeds could be possible because of local enhancements to the magnetic field. When this happens, modeling that is more accurate than EMHD becomes necessary for the final stages of the implosion. These effects are further enhanced in 2D by the generation of vortices, which propagate faster than the 1D shock.

\end{document}